\documentclass[twocolumn,showpacs,amsmath,amssymb,aps,pra,floatfix,superscriptaddress,10pt]{revtex4-2}

\usepackage[latin1]{inputenc}
\usepackage[american]{babel}
\usepackage[T1]{fontenc}
\usepackage{amsmath}
\usepackage{upgreek}
\usepackage{mhchem}
\usepackage[dvipsnames]{xcolor}
\usepackage{graphicx}
\usepackage{multirow}
\usepackage{physics}
\usepackage[caption=false]{subfig}
\usepackage{changes}
\usepackage{mathtools}
\usepackage{tabularx}
\usepackage{booktabs}
\usepackage{float}

\usepackage{todonotes}
\usepackage[colorlinks=true, linkcolor=blue, citecolor=blue, urlcolor=blue]{hyperref}

\begin{document}

\title{Tackling the uncertainty of the nuclear-polarization correction to the bound-electron $\boldsymbol{g}$ factor by means of the nuclear Skyrme interaction}

\date{\today}

\author{Arnab~Choudhury} 
\affiliation{Max-Planck-Institut f\"{u}r Kernphysik, Saupfercheckweg 1, 69117 Heidelberg, Germany}

\author{Igor~A.~Valuev}
\email[Email: ]{igor.valuev@mpi-hd.mpg.de} 
\affiliation{Max-Planck-Institut f\"{u}r Kernphysik, Saupfercheckweg 1, 69117 Heidelberg, Germany}

\begin{abstract}
The Coulomb part of the leading-order nuclear-polarization correction to the bound-electron $g$~factor of hydrogenlike ions is investigated in a microscopic approach from the nuclear point of view.
To this end, the effective Skyrme force is employed to model nucleon-nucleon interactions, with the energies and the reduced transition probabilities of collective nuclear excitations being obtained in the Hartree-Fock-based random-phase approximation.
These nuclear parameters serve as input for the nuclear-polarization correction, evaluated via effective self-energy diagrams where the photon propagator is modified by a nuclear-polarization insertion.
A diverse set of Skyrme parameterizations is probed for $^{40}\text{Ca}^{19+}$, $^{60}\text{Ni}^{27+}$, $^{90}\text{Zr}^{39+}$, and $^{120}\text{Sn}^{49+}$, and the results are compared to the common approach involving experimental nuclear data and estimates based on energy-weighted sum rules.
As a result, tighter constraints on the theoretical uncertainties of the nuclear-polarization corrections are obtained, providing key input for high-precision measurements of the bound-electron $g$~factors of heavy hydrogenlike ions.
\end{abstract}

\maketitle

\section{Introduction}
To date, the most precise verifications of quantum field theory have come from quantum electrodynamics~(QED), especially from measurements~\cite{2008_Hanneke, 2023_Fan} and calculations~\cite{2019_Aoyama, 2024_Volkov} of electron's anomalous magnetic moment.
While a free electron represents the simplest possible system resulting in the highest precision, bound-electron measurements in highly charged ions provide a unique opportunity to test QED in the strongest electromagnetic fields currently accessible for experimental investigation.
Over the last decades there has been a progression of such measurements toward higher atomic numbers up to hydrogenlike $^{118}\text{Sn}^{49+}$~\cite{2023_Morgner} while maintaining relative experimental uncertainties below $10^{-9}$.

However, the access to ever-stronger electromagnetic fields in the medium- and high-$Z$ regimes comes with the price of a higher sensitivity to nuclear effects such as finite nuclear size (FNS) and nuclear polarization (NP).
The former affects the bound-electron $g$ factor of $^{118}\text{Sn}^{49+}$ already at the level of $10^{-5}$~\cite{2023_Morgner}, and, with the recent advances in two-loop QED calculations~\cite{2025_Sikora}, the FNS uncertainty of $\sim 10^{-8}$ is now comparable to that of the bound QED theory.
As for the NP correction, although its magnitude is significantly smaller, approaching only about $10^{-8}$ for $Z=50$, the corresponding uncertainties are very large and typically estimated to be 30\nobreakdash--50\%.

On the other hand, instead of seeing nuclear effects as an obstacle, one could turn it around and use $g$ factors of heavy hydrogenlike ions to extract nuclear charge radii or potentially even probe internal nuclear dynamics.
This is of particular interest due to the fact that radius extraction from muonic-atom measurements has been facing challenges and discrepancies in heavy systems~\cite{1985_Phan, 1988_Bergem, 1990_Piller, 2025_Sun, 2025_Beyer}.
In this regard, there is an exciting possibility to bridge together muonic-atom spectroscopy and high-precision measurements of electronic systems in order to gain new insights into nuclear properties.
At the same time, such investigations in the medium- and high\nobreakdash-$Z$ regimes would also face limitations from the large uncertainties in the NP correction, which takes into account all the intricate details of the dynamic nuclear structure beyond the nuclear size described by a static charge distribution.
As the NP effect requires information about the entire nuclear excitation spectrum, it is extremely challenging to assess the accuracy of NP calculations.

In this paper, we address this challenge and present a thorough uncertainty analysis for the NP correction to the bound-electron $g$ factor by employing microscopic nuclear models based on the Skyrme interaction~\cite{skyrme_rpa}.
For this purpose, we obtain the energies and the reduced transition probabilities of excited nuclear states by means of the Hartree-Fock-based random-phase approximation.
This significantly extends the commonly used nuclear input for NP calculations based only on available experimental parameters and energy-weighted sum rules~\cite{1978_Rinker, 1996_Nefiodov}.
By studying wide ranges of Skyrme parameterizations for $^{40}\text{Ca}^{19+}$, $^{60}\text{Ni}^{27+}$, $^{90}\text{Zr}^{39+}$, and $^{120}\text{Sn}^{49+}$, we believe to have reduced the uncertainty of the Coulomb part of the leading-order NP correction down to 10\nobreakdash--12\% for $^{60}\text{Ni}^{27+}$, $^{90}\text{Zr}^{39+}$, and $^{120}\text{Sn}^{49+}$ and 17\% for $^{40}\text{Ca}^{19+}$, which represents an important step forward in enhancing the potential applications of the $g$-factor measurements of heavy hydrogenlike ions.

The relativistic system of units ($\hbar = c =1$) as well as the Heaviside-Lorentz units ($\epsilon_0=\mu_0=1$) are used in this paper such that the fine-structure constant ($\alpha$) is related to the elementary charge ($e$) as $\alpha=e^2/4\pi$.
Three-vectors are represented by upright bold letters with the usual notation $\textbf{r} = (x,y,z)$ everywhere except for Subsection~\ref{NP_insertion}, where regular typeface is reserved for four-vectors~($x$), while the magnitudes of the corresponding three-vectors~($\textbf{x}$) are denoted by regular upright letters~($\mathrm{x}$).

\section{Formalism}

\subsection{Bound-electron $\boldsymbol{g}$ factor}

In simplest terms, a $g$ factor can be defined as a constant of proportionality relating the magnetic moment $\hat{\boldsymbol{\mu}}$ (in units of the Bohr magneton $\mu_\text{B}=\frac{e}{2m_\text{e}}$) and the angular momentum $\hat{\textbf{M}}$ of an electron:
\begin{equation}
    \frac{\hat{\boldsymbol{\mu}}}{\mu_\text{B}}=-g \hat{\textbf{M}}.
    \label{eq:gfactor_1}
\end{equation}
The magnetic moment of a bound electron arises from both its spin and its orbital angular momentum due to the motion in an external potential. 
The total magnetic moment of a bound electron is given by
\begin{equation}
    \frac{\hat{\boldsymbol{\mu}}_{\text{tot}}}{\mu_\text{B}}=-(g_{\mathrm{S}}\hat{\textbf{S}} + g_{\mathrm{L}} \hat{\textbf{L}}),
    \label{eq:total_gfactor}
\end{equation}
where $\hat{\textbf{S}}$ and $\hat{\textbf{L}}$ are the spin and the orbital components of the angular momentum, respectively, with the corresponding $g$ factors $g_{\mathrm{S}}$ and $g_{\mathrm{L}}$.

Since the total magnetic moment is not collinear with the total angular momentum 
$\hat{\textbf{J}}=\hat{\textbf{L}}+\hat{\textbf{S}}$, one cannot, strictly speaking, define a $g$ factor $g_{\mathrm{J}}$ corresponding to $\hat{\textbf{J}}$ in the sense of Eq.~\eqref{eq:gfactor_1}. 
However, it can be shown that, due to the projection theorem~\cite{1957_Rose}, this proportionality still holds for $\hat{\textbf{J}}$ in terms of expectations values:
\begin{equation}
\frac{\bra{jm}\hat{\boldsymbol{\mu}}_{\text{tot}}\ket{jm}}{\mu_\text{B}}=-g_{\mathrm{J}}\bra{jm}\hat{\textbf{J}}\ket{jm},
\label{eq:Expectation_Value}
\end{equation}
where $\ket{jm}$ denotes the eigenbasis for the total angular momentum and its $z$ component.
In other words, the component of $\hat{\boldsymbol{\mu}}_{\text{tot}}$ perpendicular to
$\hat{\textbf{J}}$ does not contribute to the expectation value $\bra{jm}\hat{\boldsymbol{\mu}}_{\text{tot}}\ket{jm}$, which makes $\hat{\boldsymbol{\mu}}_{\text{tot}}$ \textit{effectively} lie in the direction of $\hat{\textbf{J}}$.
The coefficient $g_{\mathrm{J}}$ is called the Land\'e $g$ factor, and Eq.~\eqref{eq:Expectation_Value} allows one to describe the Zeeman splitting of atomic energy levels in a weak external magnetic field.
This can be immediately seen for the interaction term ${\hat{H}_{\text{mag}}}=-\hat{\boldsymbol{\mu}}_{\text{tot}}\cdot\textbf{B}$ from the non-relativistic Pauli equation, in which case the energy shift in a magnetic field $\textbf{B}=(0,0,B)$ along the $z$ direction is given by
\begin{equation}
    \Delta E=\bra{jm}-\hat{\boldsymbol{\mu}}_{\text{tot}}\cdot{\textbf{B}}\ket{jm} = g_{\mathrm{J}}\mu_\text{B} Bm.
\label{eq:gfactor_2}
\end{equation}

Finally, in the relativistic case, a more general definition of a $g$ factor can be obtained from considering the stationary Dirac equation:
\begin{equation}
\label{eq:Dirac}
    \left[ -i\nabla\cdot\boldsymbol{\alpha} +\beta m_{\text{e}} + V(\textbf{r}) \right]\psi(\textbf{r}) = \varepsilon\psi(\textbf{r}),
\end{equation}
where $\boldsymbol{\alpha}$ and $\beta$ are the Dirac matrices, and $V(\textbf{r})$ is the electrostatic potential of a nucleus.
For an arbitrary central potential $V(r)$ the four-component eigenfunctions~$\psi(\textbf{r})$ split into radial and angular parts as
\begin{equation}
    \psi_{n\kappa m}(\textbf{r}) = \frac{1}{r}
    \begin{pmatrix}
        G_{n\kappa}(r) \mathit{\Omega}_{\kappa m}(\Omega_{\textbf{r}})\\
        i F_{n\kappa}(r) \mathit{\Omega}_{-\kappa m}(\Omega_{\textbf{r}})
    \end{pmatrix},
    \label{eq:solution}
\end{equation}
where $n$ is the principal quantum number, and $\kappa$ is the relativistic angular momentum quantum number. 
The spherical spinors $\mathit{\Omega}_{\pm \kappa m}(\Omega_{\textbf{r}})$ are the same for any central potential, with $\Omega_{\textbf{r}}$ denoting the angular coordinates of the solution.

\begin{figure}
\centering
 \includegraphics[angle=90]{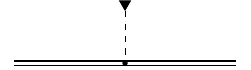}
 \caption{Feynman diagram for a bound electron (double line) interacting with an external magnetic potential (triangle).}
 \label{fig:Fig_1}
\end{figure}

In this formalism, the minimal coupling prescription leads to the magnetic interaction potential $\delta V(\textbf{r})$ given by
\begin{equation}
    \delta V(\textbf{r}) = e\boldsymbol{\alpha}\cdot \textbf{A}(\textbf{r}),
    \label{eq:magnetic_pot}
\end{equation}
where the vector potential can be chosen in the form $\textbf{A}(\textbf{r}) = [{\textbf{B}}\times{\textbf{r}}]/2$ such that the first-order energy shift evaluates to 
\begin{align}
    \Delta E_{n\kappa m}^{(1)} & = \frac{e}{2} \bra{n\kappa m}\boldsymbol{\alpha}\cdot[{\textbf{B}}\times{\textbf{r}}]\ket{n\kappa m} \notag \\
    & = \frac{e}{2} B \bra{n\kappa m}[{\textbf{r}}\times \boldsymbol{\alpha}]_z\ket{n\kappa m}.
    \label{eq:First_order}
\end{align}
It can be shown that, after evaluating the angular part of this matrix element~\cite{1961_Rose}, the energy shift reduces to the same form as in Eq.~\eqref{eq:gfactor_2}:
\begin{equation}
    \Delta E_{n\kappa m}^{(1)} = g^{(1)} \mu_{\text{B}}Bm,
    \label{eq:gfactor_3}
\end{equation}
with
\begin{equation}
    g^{(1)} = \frac{2\kappa m_\text{e}}{j(j+1)}\int_0^\infty rG_{n\kappa}(r)F_{n\kappa}(r)dr.
    \label{eq:gfactor_4}
\end{equation}
In this paper, the term bound-electron $g$ factor will be considered henceforth in the generalized sense given by Eq.~\eqref{eq:gfactor_3} as a measure of the
response to an external magnetic field.
The corresponding Feynman diagram for the interaction with the potential $\delta V$ is shown in Fig.~\ref{fig:Fig_1}.

\subsection{Nuclear-polarization insertion} \label{NP_insertion}

The NP effect is incorporated into the QED framework by expressing the total nuclear four-current density as the
following sum~\cite{1989_Plunien, 1991_Plunien, 2024_Valuev}:
\begin{equation}
    \hat{J}_{\text{tot}}(x) = J_{\text{stat}}(\textbf{x}) + \hat{J}_{\text{fluc}}(x),
    \label{eq:Nuclear_Current}
\end{equation}
where $J_{\text{stat}}$ is the classical static current corresponding to the average over the nuclear ground state, and the remaining fluctuating part $\hat{J}_{\text{fluc}}$ is the perturbation arising due to intrinsic nuclear dynamics. 
The latter is treated by associating with it an additional photon field $\hat{A}^{\mu}_{\text{fluc}}$ such that the total electromagnetic field reads
\begin{equation}
    \hat{A}^{\mu}_{\text{tot}}(x) = \hat{A}^{\mu}_{\text{free}}(x) + A^{\mu}_{\text{stat}}(\textbf{x}) + \hat{A}^{\mu}_{\text{fluc}}(x),
    \label{eq:EM_field}
\end{equation}
where $\hat{A}^{\mu}_{\text{free}}$ is the free electromagnetic field, and $A^{\mu}_{\text{stat}}$ is the classical static field generated by the nucleus, which corresponds to the potential $V$ in the Dirac equation~\eqref{eq:Dirac}.
The static part is taken into account non-perturbatively in the Furry picture resulting in a dressed electron propagator, whereas the $\hat{A}^{\mu}_{\text{fluc}}$ term can be grouped with the free field $\hat{A}^{\mu}_{\text{free}}$ defining the total quantum radiation field
\begin{equation}
    \hat{A}^{\mu}_{\text{rad}}(x) =\hat{A}^{\mu}_{\text{free}}(x)+\hat{A}^{\mu}_{\text{fluc}}(x).
    \label{eq:Radiative field}
\end{equation}
As a result, the photon propagator is modified to have the form 
\begin{align}
    \mathcal{D}_{\mu\nu}(x,x') & = -i \bra{0} T[ \hat{A}_{\mu}^{\text{rad}}(x)  \hat{A}_{\nu}^{\text{rad}}(x')] \ket{0} \notag \\
    & = D_{\mu\nu}^{\text{free}}(x-x') + D^{\text{NP}}_{\mu\nu}(x,x'),
    \label{eq:Modified propagator}
\end{align}
where $D_{\mu\nu}^{\text{free}}$ is the free photon propagator, and $D_{\mu\nu}^{\text{NP}}$ is the correction due to NP.

\begin{figure}[!t]
\centering
 \includegraphics[angle=0]{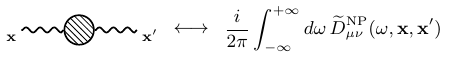}
 \caption{Feynman rule for the nuclear-polarization correction to the photon propagator. The wavy line represents the photon while the shaded circle indicates the nuclear-polarization insertion.}
 \label{fig:Fig_2}
\end{figure}

In such a framework, the only addition to the standard QED Feynman rules is that now every photon line has the corresponding NP term, which is depicted diagrammatically in Fig.~\ref{fig:Fig_2}.
The shaded circle is called the NP insertion, which contains all the information about the nuclear dynamics.
In practical calculations, e.g., using the two-time Green's function approach~\cite{2002_Shabaev}, one deals with the Fourier-transformed version of the total photon propagator with respect to time.
This is possible because the $D_{\mu\nu}^{\text{NP}}$ term can be easily shown to be homogeneous in time~\cite{1991_Plunien}. 
Similar to the standard rule for an internal photon line, the energy coordinate $\omega$ is integrated over.

\begin{figure}[!ht]
\centering
\subfloat[\centering]{{\includegraphics[]{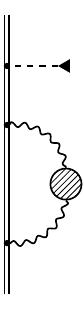} }}%
\hspace{1cm}
\subfloat[\centering]{{\includegraphics[]{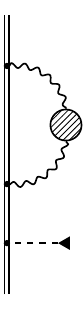} }}%
\hspace{1cm}
\subfloat[\centering]{{\includegraphics[]{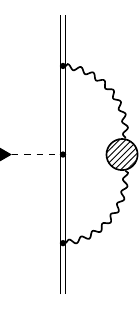} }}%
\caption{One-loop effective self-energy diagrams for a bound electron interacting with an external magnetic field.}
\label{fig:Fig_3}
\end{figure}

In this paper, the NP effect is calculated in the Coulomb approximation, where only the (0,0) component of $D_{\mu\nu}^{\text{NP}}$ is taken into account, which can be reduced to the following form~\cite{1989_Plunien, 1991_Plunien, 1996_Nefiodov}:
\begin{align}
    \notag \widetilde{D}^{\text{NP}}_{00}(\omega, \textbf{x},\textbf{x}')= & \sum_{J,M,N}F_J(\text{x})F_{J}(\text{x}')Y_{JM}(\Omega_\textbf{x})Y^*_{JM}(\Omega_{\textbf{x}'})\\
    &\times\frac{2\omega_{JN}}{\omega^2 -\omega_{JN}^2+i0^+}B(EJ)_N,
    \label{eq:Coulomb approx}
\end{align}
where the sum runs over the entire nuclear excitation spectrum such that $N$ enumerates excitations with a given total angular momentum $J$, and $M$ stands for the corresponding magnetic quantum number. 
Here, $Y_{JM}(\Omega_\textbf{x})$ are the spherical harmonics, $\omega_{JN}$ are nuclear excitation energies, $B(EJ)_N$ are reduced transition probabilities (for $J \rightarrow 0$), and $F_J(\text{x})$ are model-dependent radial functions.
In the case of electronic systems, it has been shown that it is sufficient to use the following simple radial dependence~\cite{2024_Valuev}:
\begin{align}
&\begin{aligned}
\label{eq:F_0}
F_0(\mathrm{x}) & = \dfrac{5\sqrt{\pi}}{2R_0^3} \left[ 1 - \left( \dfrac{\mathrm{x}}{R_0} \right)^2 \right] \theta(R_0 - \mathrm{x}), 
\end{aligned}\\
&\begin{aligned}
\label{eq:F_L}
F_J(\mathrm{x}) = \dfrac{4\pi}{(2J+1)R_0^J} \biggl[ &\dfrac{\mathrm{x}^J}{R_0^{J+1}} \theta(R_0-\mathrm{x}) \\
+ &\dfrac{R_0^J}{\mathrm{x}^{J+1}} \theta(\mathrm{x}-R_0) \biggr], \quad J \geq 1,
\end{aligned}
\end{align}
with $R_0$ being the radius of the nucleus taken as a homogeneously charged sphere.

\subsection{Leading-order NP corrections to the $\boldsymbol{g}$ factor}

Having introduced the notion of the modified photon propagator, one is immediately led to the leading-order NP corrections to the bound-electron $g$ factor as the effective self-energy diagrams shown in Fig.~\ref{fig:Fig_3}.
It should be noted that similar diagrams involving vacuum polarization do not need to be taken into account since they have been shown to be effectively absorbed in the FNS correction~\cite{2024_Valuev}.

The application of the two-time Green's function formalism here is analogous to the case of the corresponding diagrams without the NP insertions, which has been described in detail in Ref.~\cite{2002_Shabaev}.
In particular, the energy shift due to the diagrams in Fig.~\ref{fig:Fig_3}(a,b) is split into two contributions, the so-called irreducible and reducible parts.
Since the expressions for these two diagrams are complex conjugates of each other, one simply needs to multiply the real-valued contributions from the diagram in Fig.~\ref{fig:Fig_3}(a) by a factor of 2.
Similar to Ref.~\cite{2002_Nefiodov}, we obtain the following expressions for the irreducible and reducible energy shifts for a state $\ket{i}=\ket{n_i \kappa_i m_i}$:

\begin{widetext}
\begin{align}
&\begin{aligned}
 \label{eq:g_irr}
    \Delta E_{i,\text{irr}}^{(2,\text{a}+\text{b})} =  \alpha m_{\text{e}}\mu_{\text{B}}Bm_i \frac{2\kappa_i}{j_i(j_i+1)(2j_i+1)}\,& \sum_{J,N} (2J+1) B(EJ)_N\sum^{n_1\neq n_i}_{n_{1},n_{2},\kappa_{2}} [C_{J}(\kappa_i,\kappa_{2})]^2
    \\
    & \times \frac{\bra{n_i\kappa_i}r\sigma_x \ket{n_{1}\kappa_{i}}\bra{n_{1}\kappa_{i}}F_{J}\mathbb{I} \ket{n_{2}\kappa_{2}}\bra{n_2\kappa_2} F_{J}\mathbb{I} \ket{n_{i}\kappa_{i}}}{[\varepsilon_{n_i\kappa_i}-\varepsilon_{n_{1}\kappa_i}][\varepsilon_{n_i\kappa_i}-\varepsilon_{n_{2}\kappa_{2}}-\mathrm{sgn}(\varepsilon_{n_{2}\kappa_{2}})\omega_{JN}]},
\end{aligned} \\[15pt]
&\begin{aligned}
 \label{eq:g_red}
    \Delta E_{i,\text{red}}^{(2,\text{a}+\text{b})} = -\alpha m_{\text{e}}\mu_{\text{B}}Bm_i \frac{\kappa_i}{j_i(j_i+1)(2j_i+1)} & \bra{n_i\kappa_i}r\sigma_x\ket{n_i\kappa_i} \sum_{J,N}(2J+1)B(EJ)_N
    \\
    \times & \sum_{n_{1},\kappa_{1}}[C_{J}(\kappa_i,\kappa_{1})]^2\frac{\bra{n_i\kappa_i}F_{J}\mathbb{I}\ket{n_{1}\kappa_{1}}^2 }{[\varepsilon_{n_i\kappa_i}-\varepsilon_{n_1\kappa_1}-\mathrm{sgn}(\varepsilon_{n_1\kappa_1})\omega_{JN}]^2},
\end{aligned}
\end{align}
\end{widetext}
with
\begin{align}
    \notag C_J(\kappa_1,\kappa_2) &= (-1)^{j_1+\frac{1}{2}}\sqrt{(2j_1+1)(2j_2+1)}\\
    &\times\Pi(l_1+l_2+J)\times \begin{pmatrix}
        j_2&J&j_1\\
        \frac{1}{2}&0&-\frac{1}{2}
    \end{pmatrix},
    \label{eq:C_J}
\end{align}
where $\Pi(k)$ is the parity operator returning 1 if $k$ is even and 0 otherwise, $l$ is the orbital momentum quantum number, and the last factor is a 3-j symbol.
In the radial matrix elements, $\sigma_x$ stands for one of the Pauli matrices, and $\mathbb{I}$ denotes the unity $2 \times 2$ matrix such that the explicit expressions are given by
\begin{align}
&\begin{aligned}
\bra{n_1\kappa_1}r\sigma_x\ket{n_2\kappa_2} = \int_{0}^{\infty} r [ & G_{n_1\kappa_1}(r) F_{n_2\kappa_2}(r) \\ 
 + & F_{n_1\kappa_1}(r) G_{n_2\kappa_2}(r) ] dr,
\end{aligned} \\
&\begin{aligned}
\bra{n_1\kappa_1}F_{J}\mathbb{I}\ket{n_2\kappa_2} = \int_{0}^{\infty} F_J(r) [ & G_{n_1\kappa_1}(r) G_{n_2\kappa_2}(r) \\
+ & F_{n_1\kappa_1}(r) F_{n_2\kappa_2}(r) ] dr.
\end{aligned}
\end{align}

The correction due to the diagram in Fig.~\ref{fig:Fig_3}(c) is also divided into two parts, the residual and the pole contributions, based on the poles in the integrals used to derive these expressions.
The corresponding energy shifts read
\begin{widetext}
\begin{align}
&\begin{aligned}
 \label{eq:g_res}
    & \Delta E_{i,\text{res}}^{(2,\text{c})} = \alpha m_{\text{e}}\mu_{\text{B}}Bm_i \frac{4}{\sqrt{j_i(j_i+1)(2j_i+1)}} \sum_{J,N} (-1)^{J}(2J+1) B(EJ)_N\sum^{n_1\neq n_2}_{n_{1},\kappa_{1},n_{2},\kappa_{2}} (-1)^{j_{2}+\frac{1}{2}} \kappa_1 \sqrt{\dfrac{2(2j_{2}+1)}{2j_{1}+1}}
    \\
    & \times \Pi (l_{1}+l_{2}) C_{J}(\kappa_i,\kappa_{1})C_{J}(\kappa_{2},\kappa_{i})
    \begin{Bmatrix}
        j_{i}&1&j_{i}\\
       j_{2}&J&j_{1}
    \end{Bmatrix} 
    \begin{pmatrix}
        j_{1}&j_{2}&1\\
        \frac{1}{2}&\frac{1}{2}&-1
    \end{pmatrix} 
    \frac{\bra{n_i\kappa_i}F_{J}\mathbb{I}\ket{n_{1}\kappa_{1}}\bra{n_{1}\kappa_{1}}r\sigma_x \ket{n_{2}\kappa_{2}}\bra{n_{2}\kappa_{2}}F_{J}\mathbb{I}\ket{n_i\kappa_i}}{[\varepsilon_{n_1\kappa_1}-\varepsilon_{n_2\kappa_2}][\varepsilon_{n_i\kappa_i}-\varepsilon_{n_1\kappa_1}-\mathrm{sgn}(\varepsilon_{n_1\kappa_1})\omega_{JN}]},
\end{aligned}\\[15pt]
&\begin{aligned}
 \label{eq:g_pol}
    \Delta E_{i,\text{pole}}^{(2,\text{c})} = \alpha m_{\text{e}}\mu_{\text{B}}Bm_i \frac{1}{\sqrt{j_i(j_i+1)(2j_i+1)}}
    \sum_{J,N} & (-1)^{J} (2J+1)B(EJ)_N\sum_{n_1,\kappa_1} \kappa_{1} \sqrt{\frac{2j_1+1}{j_1(j_1+1)}} 
    \begin{Bmatrix}
       j_{i}&1&j_{i}\\
       j_{1}&J&j_{1}
    \end{Bmatrix} 
    \\
    & \times C_{J}(\kappa_i,\kappa_{1})C_{J}(\kappa_{1},\kappa_{i}) 
    \frac{\bra{n_i\kappa_i}F_{J}\mathbb{I}\ket{n_{1}\kappa_{1}}^2\bra{n_{1}\kappa_{1}} r\sigma_x \ket{n_{1}\kappa_{1}}}{[\varepsilon_{n_i\kappa_i}-\varepsilon_{n_1\kappa_1}-\mathrm{sgn}(\varepsilon_{n_1\kappa_1})\omega_{JN}]^2},
\end{aligned}
\end{align}
\end{widetext}
where the term in braces is a 6-j symbol.

Finally, the leading-order NP correction to the bound-electron $g$ factor is simply obtained as
\begin{equation}
    g^{(2)}_{i} = \dfrac{\Delta E^{(2)}_i}{\mu_{\text{B}}Bm_i}.
\end{equation}

\subsection{Nuclear input}

The main source of uncertainty in evaluating the NP effect comes from the nuclear input needed for the correction to the photon propagator given by Eq.~\eqref{eq:Coulomb approx}.
The challenge is that experimental data for the parameters $\omega_{JN}$ and $B(EJ)_N$ are available only for a limited number of low-lying states at best, whereas a significant or even dominant contribution to the NP correction stems from high-frequency collective nuclear excitations known as the giant resonances.
Therefore, the most common approach is to combine the available experimental data for low-lying nuclear states with simple estimates for the giant resonances based on energy-weighted sum rules (EWSR)~\cite{1978_Rinker, 1996_Nefiodov}.
The latter is achieved by assuming that for each multipolarity $J$ and isospin $\tau$ the giant resonances of a nucleus with the mass number $A$ are concentrated in a single effective state with the energy
\begin{align}
\label{eq:sum_rules_energies}
\langle \omega_{J} \rangle & = [100(1-\tau)+200\tau](1-A^{-1/3})A^{-1/3}, \, J = 0, \notag \\
& = 95(1-A^{-1/3})A^{-1/3}, \, J = 1, \notag \\
& = [75(1-\tau)+160\tau](1-A^{-1/3})A^{-1/3}, \, J \geq 2.
\end{align}
The corresponding reduced transition probabilities are then obtained from the following relations:
\begin{align}
\label{eq:sum_rules_probabilities}
\langle \omega_{0}\rangle B(E0) & = \dfrac{25 Ze^{2}}{4\pi M_{\text{p}}} \langle r^{2}_{\text{p}} \rangle \left( \dfrac{Z}{A} (1-\tau) + \dfrac{N}{A} \tau \right), \notag \\
\langle \omega_{J} \rangle B(EJ) & = \dfrac{J (2J+1) Ze^{2}}{8\pi M_{\text{p}}} \langle r^{2J-2}_{\text{p}} \rangle \notag \\ 
& \times \left( \dfrac{Z}{A} (1-\tau) + \dfrac{N}{A} \tau \right), \, J \geq 1,
\end{align}
where $N$ is the neutron number, $M_{\text{p}}$ is the proton mass, and the expectation value for protons in the initial state is estimated as $\langle r^{2J-2}_{\text{p}} \rangle = 3R_{0}^{2J-2}/(2J+1)$.
We note that for $J=1$ only the isovector ($\tau=1$) resonances are considered since the isoscalar ($\tau=0$) case corresponds to the motion of the center of mass of the entire nucleus.

It is the use of such simple models and the incompleteness of the experimental data that commonly lead to assigning very conservative estimates of uncertainties to calculated NP corrections.
In this paper, we consider an additional source of nuclear input from microscopic calculations of the parameters $\omega_{JN}$ and $B(EJ)_N$.
For this purpose, we employ the \verb|skyrme_rpa| program \cite{skyrme_rpa}, which is based on the effective Skyrme nucleon-nucleon interaction.
In this program, an excitation spectrum for a given $J$ is calculated in the framework of the Hartree-Fock-based random-phase approximation.
The completeness of the obtained spectra can be readily controlled by the fulfillment of the EWSR.
Since this approach is restricted to spherical or mostly spherical nuclei with at least one closed major shell, we consider $^{40}\text{Ca}$, $^{60}\text{Ni}$, $^{90}\text{Zr}$, and $^{120}\text{Sn}$.
In our calculations we include multipolarities up to $J=5$.

\section{Results and discussion}

\begin{table}[!tbp]
\begin{center}
\caption{Different parts of the leading-order nuclear-polarization corrections to the bound-electron $g$-factor of $^{120}\text{Sn}^{49+}$ in the $1s_{1/2}$, $2s_{1/2}$, and $2p_{1/2}$ states. The results are shown for the SLy5 nuclear model. The numbers in square brackets indicate powers of 10.}
{\renewcommand{\arraystretch}{1.2}
\renewcommand{\tabcolsep}{0.29cm}
\begin{tabular}{lccc}
\hline \hline
& $1s_{1/2}$ & $2s_{1/2}$ & $2p_{1/2}$ \\
\hline
Irreducible  &  $-7.91[-9]\phantom{1}$ & $-1.11[-9]\phantom{1}$ & $-2.87[-11]$ \\
Reducible    &  $-2.06[-10]$ & $-2.99[-11]$ & $-2.87[-13]$ \\
Residual     &  $+6.97[-12]$ & $+3.95[-13]$ & $-1.06[-12]$ \\
Pole         &  $+6.57[-12]$ & $+9.04[-13]$ & $+1.96[-14]$\\[5pt]
Total        &  $-8.10[-9]\phantom{1}$ & $-1.14[-9]\phantom{1}$ & $-3.01[-11]$ \\
\hline \hline
\end{tabular}} 
\label{tab:g_cont}
\end{center}
\end{table}

\begin{table}[!tbp]
\begin{center}
\caption{Contributions from different multipolarities of nuclear excitations to the leading-order nuclear-polarization corrections to the bound-electron $g$-factor of $^{120}\text{Sn}^{49+}$ in the $1s_{1/2}$, $2s_{1/2}$, and $2p_{1/2}$ states. The results are shown for the SLy5 nuclear model. The numbers in square brackets indicate powers of 10.}
{\renewcommand{\arraystretch}{1.2}
\renewcommand{\tabcolsep}{0.39cm}
\begin{tabular}{lccc}
\hline \hline
& $1s_{1/2}$ & $2s_{1/2}$ & $2p_{1/2}$ \\
\hline
$0^{+}$  & $-6.52[-10]$ & $-9.21[-11]$ & $-2.38[-12]$ \\
$1^{-}$  & $-5.65[-9]\phantom{1}$ & $-7.97[-10]$ & $-2.12[-11]$ \\
$2^{+}$  & $-1.10[-9]\phantom{1}$ & $-1.56[-10]$ & $-3.92[-12]$ \\
$3^{-}$  & $-4.93[-10]$ & $-6.97[-11]$ & $-1.78[-12]$ \\
$4^{+}$  & $-1.31[-10]$ & $-1.85[-11]$ & $-4.76[-13]$ \\
$5^{-}$  & $-7.35[-11]$ & $-1.04[-11]$ & $-2.67[-13]$ \\[5pt]
Total    & $-8.10[-9]\phantom{1}$ & $-1.14[-9]\phantom{1}$ & $-3.01[-11]$ \\
\hline \hline
\end{tabular}} 
\label{tab:J_cont}
\end{center}
\end{table}

\begin{table*}[!htbp]
\caption{Model dependence of the total leading-order nuclear-polarization corrections to the bound-electron $g$ factor of hydrogenlike $^{120}\text{Sn}^{49+}$, $^{90}\text{Zr}^{39+}$, $^{60}\text{Ni}^{27+}$, and $^{40}\text{Ca}^{19+}$ in the $1s_{1/2}$, $2s_{1/2}$, and $2p_{1/2}$ states.
The average values and the half-range uncertainties are given for the nine Skyrme nuclear models.
The last row contains the results obtained by using experimental nuclear parameters for low-lying states and the energy-weighted-sum-rule estimates for the giant resonances.
The numbers in square brackets indicate powers of 10 for each particular ion.}
{\renewcommand{\arraystretch}{1.2}
\renewcommand{\tabcolsep}{0.21cm}
\begin{tabular*}{\textwidth}{lcccccccccccc}
\hline \hline
\multirow{2}{*}{Model} & \multicolumn{3}{c}{$^{120}\text{Sn}^{49+}[-10]$} & \multicolumn{3}{c}{$^{90}\text{Zr}^{39+}[-10]$} & \multicolumn{3}{c}{$^{60}\text{Ni}^{27+}[-11]$} & \multicolumn{3}{c}{$^{40}\text{Ca}^{19+}[-12]$} \\
\cmidrule(lr{1pt}){2-4}
\cmidrule(lr{1pt}){5-7}
\cmidrule(lr{1pt}){8-10}
\cmidrule(lr{1pt}){11-13}
& $1s_{1/2}$ & $2s_{1/2}$ & $2p_{1/2}$ & $1s_{1/2}$ & $2s_{1/2}$ & $2p_{1/2}$ & $1s_{1/2}$ & $2s_{1/2}$ & $2p_{1/2}$ & $1s_{1/2}$ & $2s_{1/2}$ & $2p_{1/2}$ \\
\hline
KDE0  & $-79.8$ & $-11.3$ & $-0.30$ & $-19.8$ & $-2.68$ & $-0.044$ & $-28.7$ & $-3.73$ & $-0.031$ & $-52.0$ & $-6.62$ & $-0.031$ \\
SKX   & $-81.0$ & $-11.4$ & $-0.30$ & $-20.6$ & $-2.78$ & $-0.046$ & $-30.3$ & $-3.93$ & $-0.033$ & $-52.2$ & $-6.65$ & $-0.031$ \\
SLy5  & $-81.0$ & $-11.4$ & $-0.30$ & $-21.0$ & $-2.84$ & $-0.047$ & $-30.7$ & $-3.98$ & $-0.033$ & $-55.7$ & $-7.09$ & $-0.033$ \\
BSk14 & $-80.5$ & $-11.4$ & $-0.30$ & $-20.8$ & $-2.80$ & $-0.046$ & $-29.8$ & $-3.86$ & $-0.032$ & $-54.6$ & $-6.95$ & $-0.033$ \\
SAMi  & $-86.3$ & $-12.2$ & $-0.32$ & $-22.2$ & $-3.00$ & $-0.050$ & $-34.4$ & $-4.46$ & $-0.037$ & $-56.8$ & $-7.24$ & $-0.034$ \\
NRAPR & $-84.9$ & $-12.0$ & $-0.31$ & $-21.8$ & $-2.95$ & $-0.049$ & $-36.1$ & $-4.68$ & $-0.039$ & $-53.6$ & $-6.83$ & $-0.032$ \\
SkP   & $-83.9$ & $-11.8$ & $-0.31$ & $-22.0$ & $-2.97$ & $-0.049$ & $-31.8$ & $-4.13$ & $-0.034$ & $-58.4$ & $-7.45$ & $-0.035$ \\
SkM*  & $-87.3$ & $-12.3$ & $-0.32$ & $-22.7$ & $-3.06$ & $-0.051$ & $-32.9$ & $-4.27$ & $-0.036$ & $-60.5$ & $-7.70$ & $-0.036$ \\
SGII  & $-90.1$ & $-12.7$ & $-0.33$ & $-22.8$ & $-3.08$ & $-0.051$ & $-33.6$ & $-4.36$ & $-0.036$ & $-58.2$ & $-7.41$ & $-0.035$ \\[5pt]
Average & $-83.9$ & $-11.8$ & $-0.31$ & $-21.5$ & $-2.91$ & $-0.048$ & $-32.0$ & $-4.16$ & $-0.035$ & $-55.8$ & $-7.10$ & $-0.033$ \\
Range/2 & $\phantom{-8}5.2$ & $\phantom{-1}0.7$ & $\phantom{-}0.02$ & $\phantom{-2}1.5$ & $\phantom{-}0.20$ & $\phantom{-}0.004$ & $\phantom{-3}3.7$ & $\phantom{-}0.48$ & $\phantom{-}0.004$ & $\phantom{-5}4.3$ & $\phantom{-}0.54$ & $\phantom{-}0.003$ \\
[5pt]
Exp.+EWSR & $-75.2$ & $-10.6$ & $-0.28$ & $-19.3$ & $-2.61$ & $-0.043$ & $-28.2$ & $-3.66$ & $-0.030$ & $-46.3$ & $-5.90$ & $-0.027$ \\
\hline \hline
\end{tabular*}} 
\label{tab:res}
\end{table*}

Before examining the nuclear model dependence, we begin by comparing the magnitudes of the irreducible, reducible, residual, and pole parts of the leading-order NP correction to the bound-electron $g$-factor in Table~\ref{tab:g_cont} using $^{120}\text{Sn}^{49+}$ and the SLy5 Skyrme parameterization~\cite{SLy5} as an example.
We present our results for the $1s_{1/2}$, $2s_{1/2}$, and $2p_{1/2}$ states, and the corresponding total NP corrections decrease by around an order of magnitude for the next principal or angular quantum number. 
It can be immediately seen that the diagrams in Fig.~\ref{fig:Fig_3}(a,b) are responsible for most of the effect, in particular the irreducible part.
While the correction from the diagram in Fig.~\ref{fig:Fig_3}(c) contributes with the opposite sign in the case of the $1s_{1/2}$ and $2s_{1/2}$ states, its magnitude is virtually negligible.
It should be noted that the net sign of this contribution and the hierarchy of the residual and the pole parts depends on the ion and the electronic state under consideration.
For example, it is negative in the case of the $2p_{1/2}$ state of $^{120}\text{Sn}^{49+}$, and the magnitude of the residual part turns out to be larger than that of the reducible one.
However, even in this case most of the NP effect is still captured by the irreducible contribution.
In practical terms, this means that a reliable estimate of the leading-order NP correction can be obtained by considering only the diagrams in Fig.~\ref{fig:Fig_3}(a,b), which are easier and computationally less expensive to calculate than the diagram in Fig.~\ref{fig:Fig_3}(c).

Next, for the same example of $^{120}\text{Sn}^{49+}$ and the SLy5 nuclear model, we show in Table~\ref{tab:J_cont} the contributions from different multipolarities $J$ (indicated with the corresponding natural parities as superscripts) of nuclear excited states entering the summations in Eqs.~\eqref{eq:g_irr}, \eqref{eq:g_red}, \eqref{eq:g_res}, and \eqref{eq:g_pol}.
The dominant contribution comes from $1^{-}$ excitations, more specifically from the giant dipole resonances.
However, as discussed in Ref.~\cite{2024_Valuev}, this is characteristic of spherical nuclei, whereas for deformed nuclei low-lying quadrupole excitations are often responsible for the largest contributions to the NP corrections.
We also note the slow convergence with respect to $J$ such that the commonly neglected $4^{+}$ and $5^{-}$ contributions still add up to about 2.5\% of the total values according to our calculations.

Finally, in Table~\ref{tab:res} we present the total leading-order NP corrections to the bound-electron $g$ factor of four hydrogenlike ions $^{120}\text{Sn}^{49+}$, $^{90}\text{Zr}^{39+}$, $^{60}\text{Ni}^{27+}$, and $^{40}\text{Ca}^{19+}$ in the $1s_{1/2}$, $2s_{1/2}$, and $2p_{1/2}$ states.
The theoretical uncertainty comes primarily from the choice of a nuclear model used to generate the parameters $\omega_{JN}$ and $B(EJ)_N$.
In order to quantify this model dependence, we performed microscopic calculations for the same set of nine Skyrme parameterizations~\cite{SLy5, KDE0, SKX, BSk14, SAMi, NRAPR, SkP, SkM, SGII} that were used in Ref.~\cite{2022_Valuev} and shown to be a good representation of all realistic models based on rather conservative constraints on various nuclear saturation properties.
These Skyrme values are compared to the NP corrections obtained by using the combination of experimental nuclear input for low-lying states~\cite{Nucl_data_40, Nucl_data_60, Nucl_data_90, Nucl_data_120} and the EWSR\nobreakdash-based estimates for the giant resonances.
To facilitate this comparison, we also provide the averages and the half-range uncertainties for the Skyrme results.

The first immediate observation from Table~\ref{tab:res} is that for all the ions the ``Exp.+EWSR'' values are smaller than any of the Skyrme ones. 
This is rather expected due to the incompleteness of the available experimental input for low-lying nuclear states; however, it is reasonable to assume that enough of the most prominent nuclear excitations are still captured.
In fact, we find that the ``Exp.+EWSR'' values agree with the average Skyrme results within only two half-range uncertainties for $^{120}\text{Sn}^{49+}$, $^{90}\text{Zr}^{39+}$, and $^{60}\text{Ni}^{27+}$ and within three half-range uncertainties for $^{40}\text{Ca}^{19+}$.
This agreement is rather remarkable if one also considers the enormous difference in complexity of taking into account the giant resonances in the two approaches.
While the EWSR\nobreakdash-based approximation given by Eqs.~\eqref{eq:sum_rules_energies}~and~\eqref{eq:sum_rules_probabilities} concentrates the giant resonances only in a few effective states, the microscopic Skyrme spectra are highly fragmented and consist of hundreds or even thousands of excitations.
This suggests that the giant-resonance contribution to the NP effect primarily depends on the integral properties of a nuclear spectrum rather than its fine details, which provides more confidence in the simple EWSR\nobreakdash-based estimates when microscopic calculations are not available.

The main value of the information on the model dependence in Table~\ref{tab:res} is that it can be used as a basis for more informed estimates of the theoretical uncertainties of the NP corrections.
For example, rather conservative absolute uncertainties can be obtained as the differences between the average Skyrme values and the ``Exp.+EWSR'' estimates.
Using the average Skyrme results as the central values, this choice accounts for the entire range of likely under- and overestimation.
As a result, the corresponding relative uncertainties amount to only about 10\nobreakdash--12\% for $^{120}\text{Sn}^{49+}$, $^{90}\text{Zr}^{39+}$, and $^{60}\text{Ni}^{27+}$ and 17\% for $^{40}\text{Ca}^{19+}$.

\section{Conclusions and outlook}

In this work we have presented an extensive study of the nuclear model dependence of the leading-order NP correction to the bound-electron $g$ factor by means of the Skyrme interaction.
We have also compared our microscopic calculations with the common recipe of using the combination of experimental nuclear data and EWSR\nobreakdash-based estimates, and we have found a remarkable agreement between the two approaches.
Consequently, more rigorously justified and narrower theoretical uncertainties can now be provided for the NP corrections, reducing them by up to several times compared to traditionally assumed values.
It should be mentioned that only the Coulomb part of the NP correction was considered in our calculations, whereas the transverse contribution remains a topic of ongoing research.
While the transverse part is generally considered to be negligible due to the non-relativistic nature of nuclear dynamics, it is desirable to check this assumption in light of the existing counterintuitive findings in the literature on the subject~\cite{Haga_e_2002, 2024_Valuev}.

Beyond the specific nuclei considered in this work, our results also suggest a more general confidence in the simple approach of using experimental nuclear parameters and EWSR\nobreakdash-based approximations, e.g., in the case of deformed nuclei, where more detailed microscopic calculations are not readily available.
Thus, the present analysis provides a valuable benchmark for more reliable and tighter
constraints on theoretical uncertainties of the NP corrections to the bound-electron $g$ factor across a large portion of the nuclear chart.

\begin{acknowledgments}
Arnab Choudhury thanks C.~H. Keitel for financial support of his summer internship at Max Planck Institute for Nuclear Physics in Heidelberg, Germany.
\end{acknowledgments}

\end{document}